\documentclass[aps,pra,twocolumn,halfspacing,10pt,nofootinbib]{revtex4-1}
\usepackage{epsfig}
\usepackage[colorlinks]{hyperref}
\usepackage[utf8]{inputenc}
\usepackage[english]{babel}
\usepackage{graphicx}
\usepackage{color}
\usepackage{bm}
\usepackage{amssymb}
\usepackage[intlimits]{amsmath}
\usepackage{amsmath}
\usepackage[para,online,flushleft]{threeparttable}

\begin{document}
\title{Calculation of the energy levels and hyperfine structure for Xe~II, Rn~II, and Og~II ions.}
\author{T. H. Dinh$^{1}$}
\author{V. A. Dzuba$^2$}
\author{V. V. Flambaum$^{2}$}
\affiliation{$^1$Department of Physics, Ho Chi Minh City University of education, Ho Chi Minh City 72759, Vietnam}
\affiliation{$^2$School of Physics, University of New South Wales, Sydney 2052, Australia}

\date{\today}

\begin{abstract}

Energy levels, Land\'e $g$-factors, and hyperfine-structure constants are calculated for the singly ionized noble-gas atoms Xe II, Rn II, and Og II. The calculations are performed using the configuration-interaction method with perturbative treatment of high-lying configurations. Core polarization effects in the hyperfine interaction are included within the time-dependent Hartree-Fock method. Calculations for Xe II are used to test the accuracy of the approach by comparison with available experimental data. The agreement is at the level of about one percent for the energies and typically about ten percent for the hyperfine constants, with better accuracy for states with large hyperfine constants. Predictions are then presented for Rn II and Og II, for which experimental spectroscopic data are limited or absent. Special attention is paid to Breit and quantum-electrodynamic corrections to the hyperfine structure in heavy many-electron ions. We show that these corrections may be strongly enhanced by configuration mixing when interacting states with very different hyperfine matrix elements are separated by small energy intervals. This effect is demonstrated explicitly for odd-parity $J=1/2$ states of Og II. The calculated hyperfine-structure constants for Rn II and Og II provide electronic factors needed for extracting nuclear magnetic dipole and electric quadrupole moments from future spectroscopic measurements. These results may be useful for experimental studies of radon and oganesson ions and for testing nuclear models in the superheavy region.

\end{abstract}

\maketitle

\section{Introduction}

The theoretical study of the electronic structure of superheavy elements (SHEs) provides valuable insights into the interplay between strong relativistic and correlation effects and yields predictions for the spectroscopic properties of SHEs, for which experimental data are often limited (see, e.g., recent reviews~\cite{smits2023,smits2024,ackermann2024,ye2025}). Superheavy elements are also of particular interest in the search for the predicted “island of stability,” where nuclear shell models suggest the existence of relatively long-lived isotopes~\cite{ackermann2024,DFW17-IS,Pasteka}. However, despite significant advances in the production of SHEs, direct experimental studies remain extremely challenging because of the short half-lives and low production rates of these elements.

Studies of hyperfine structure (HFS) are motivated by the possibility of extracting nuclear electromagnetic moments from spectroscopic measurements. Such information would improve our understanding of the nuclear structure of superheavy nuclei and provide valuable input for the search for the predicted island of stability. In addition, hyperfine-structure measurements can provide important benchmarks for atomic-structure calculations in the regime of strong relativistic effects, thereby improving the reliability of theoretical predictions for superheavy elements.

Currently, no experimental data are available for the spectroscopic properties of superheavy elements with $Z>103$, $Z$ is atomic number. Numerous theoretical studies have attempted to address this lack of experimental information~\cite{smits2023,smits2024}. However, some systems have still received little attention. The electronic structure of neutral oganesson (Og, $Z=118$) was investigated theoretically in our previous work~\cite{E118}. On the other hand, to the best of our knowledge, the spectroscopic properties of the singly ionized species Og~II and Rn~II have not been studied previously.

Working with ions offers several advantages. They can be confined in Paul or Penning traps~\cite{Blaum2010,deGroote2024,Werth} and storage rings~\cite{ring1,ring2}.
Mass spectrometers can be used for isotope separation. In addition, both Og~II and Rn~II possess hyperfine structure in their ground states, providing additional opportunities for spectroscopic measurements and for the extraction of nuclear electromagnetic moments.

It also turns out that the accuracy of theoretical calculations is generally higher for ions than for neutral atoms, which is beneficial for the reliable interpretation of future measurements. Two major factors determine the accuracy of calculations for many-electron systems: the number of valence electrons and the number of holes in the outermost electron shell. The smaller each of these parameters, the better accuracy can be achieved. The most favourable case is a system with a single valence electron outside closed shells, such as Rb, Cs,  Fr or similar atoms or ions~\cite{DzuFlaSilSus87,DzuFlaSus89,RbPNC12,DzuJohSaf05}.
In the case of oganesson, this would correspond to highly charged ions such as Og~VI or Og~VIII. However, the production, trapping, and spectroscopic investigation of such highly charged ions may be experimentally challenging.

Singly ionized systems such as Rn~II and Og~II represent a good compromise. Their ground-state configurations contain only one hole in the outermost shell (and typically two holes in the low-lying excited states), while the number of valence electrons is seven if the outermost $ns^2$ subshell is included in the valence space. Although seven valence electrons still constitute a challenging problem for configuration-interaction (CI) calculations, an efficient extension of the CI method, specifically developed for systems with many valence electrons, is available. This method, known as configuration interaction with perturbation theory (CIPT)~\cite{CIPT}, has been successfully applied to a variety of heavy atoms and ions with open electronic shells~\cite{E118,SHE6d,ErFm,acinides,OsIr2023,LvTs-ions}. In this paper we use the method for Xe~II, Rn~II and Og~II.

Xenon and its ions have been studied extensively, both experimentally and theoretically, and therefore provide an excellent testing ground for theoretical methods. There are several experimental~\cite{Borchers1987Rn} and theoretical~\cite{Singh2013,Salah2024} studies of neutral radon. However, to the best of our knowledge, no data on the spectroscopic properties of radon ions are currently available.

The situation is even more severe for oganesson. No experimental spectroscopic data exist for either the neutral atom or its ions. Although several theoretical studies of neutral Og have been reported~\cite{Jerabek2018,E118,Zhang2022,Kumar2021}, corresponding data for oganesson ions are lacking. The present work aims to fill this gap by providing predictions for the spectroscopic properties of Rn~II and Og~II.

\section{Method of calculation}
	
\subsection{Calculation of energy levels and $g$-factors}

In the present paper we mainly follow our previous paper on the HFS of Dy, Ho, Cf and Es~\cite{Dyhfs}. 

The ions considered in the present work have an open outer shell with the configuration $ns^2np^5$ ($n=5,6,$ and 7 for Xe~II, Rn~II, and Og~II, respectively). The spectrum of Xe~II contains low-lying states involving excitation of a $5s$ electron. This implies that the $5s$ electrons in Xe~II must be treated as valence electrons. The same is true for Rn~II and Og~II. As a result, the number of valence electrons is seven, which makes conventional configuration interaction (CI) calculations computationally demanding. Therefore, we employ the CIPT (configuration interaction with perturbation theory) method~\cite{CIPT}, which was specifically developed for systems with a large number of valence electrons. The method reduces the size of the CI matrix by neglecting the off-diagonal matrix elements between high-energy states  and treating the contributions of these states to the matrix elements between low-energy states in a perturbative manner\footnote{Contribution of the  off-diagonal matrix elements between high-energy states  to the energies of  low energy states is relatively small since in  the perturbation theory treatment it appears in the third order while direct matrix elements  between the low energy and high energy states appear in the second order ~\cite{CIPT}.} . As a consequence, the dimension of the resulting CI matrix is equal to the number of low-energy states included explicitly.

{ 
The CIPT method assumes that the many-electron wave function is dominated by a relatively small number of low-energy Slater determinants. This situation is naturally achieved when the single-electron basis states are calculated in the $V^{N-1}$ approximation, where $N$ is the total number of electrons in the ion. In this approximation, the self-consistent HF potential is obtained by performing Hartree-Fock calculations with one $p$ electron removed from the uppermost subshell of the ground-state configuration of the ion. All excited single-electron states are then calculated in this $V^{N-1}$ potential. This provides a physically realistic picture in which an excited electron moves in the field of the remaining $N-1$ atomic electrons.

The single-electron relativistic Hartree-Fock (RHF) Hamiltonian $\hat{H}^{\mathrm{RHF}}_i$ has the form
\begin{eqnarray} \label{e:RHF}
\hat{H}^{\mathrm{RHF}}_i &=& c\bm{\alpha}i\cdot{\bf \hat p}_i + (\beta_i-1)mc^2 + V_{\rm nuc}(r_i) \\
&+&V^{N-1}(r_i)+V_{\rm Breit}(r_i)+V_{\rm QED}(r_i). \nonumber
\end{eqnarray}
Here, $c$ is the speed of light, $\bm{\alpha}i$ and $\beta_i$ are the Dirac matrices, ${\bf \hat p}i$ is the momentum operator of the $i$th electron, and $m$ is the electron mass. The nuclear potential $V_{\rm nuc}(r)$ is obtained by integrating the Fermi distribution of the nuclear charge density. 
The potential $V^{N-1}(r)$ is the self-consistent RHF potential formed by $N-1$ atomic electrons.
The potentials $V_{\rm Breit}$ and $V_{\rm QED}$ represent the Breit and quantum electrodynamic (QED) corrections, respectively. The latter are discussed in more detail in Sec.~\ref{s:mixing}.

After the self-consistent relativistic Hartree-Fock procedure has converged, the Hamiltonian in Eq.~(\ref{e}) is used to generate a set of single-electron basis states using the B-spline technique~\cite{B-splines}. We employ 40 B-splines of order 9 in a cavity of radius $R_{\rm max}=40a_B$, where $a_B$ is the Bohr radius. The maximum orbital angular momentum included is $l_{\rm max}=4$, and the maximum principal quantum number is $n_{\rm max}=20$. The many-electron basis states are then constructed from these single-electron orbitals in the form of Slater determinants.

In the CIPT calculations, the seven uppermost $s$ and $p$ electrons are treated as valence electrons, while all remaining electrons constitute the core. 
The CI Hamiltonian is given by
\begin{equation}\label{e}
\hat{H}^{\mathrm{CI}}=\sum_{i=1}^{N_v}\hat{H}_i^{\mathrm{RHF-CI}}+\sum_{i<j}^{N_v}\frac{e^{2}}{|r_i-r_j|},
\end{equation}
where $i$ and $j$ label the valence electrons, $N_v=7$ is the number of valence electrons, $e$ is the electron charge, and $r_i$ is the distance of the $i$th electron from the nucleus.

The RHF operator $\hat{H}^{\mathrm{RHF-CI}}$ differs from the operator in Eq.~(\ref{e}) in that only the core electrons contribute to the electronic potentials $V$ and $V_{\rm Breit}$. Thus, $V^{N-1}$ in Eq.~(\ref{e:RHF}) is replaced by the potential $V_{\rm core}$ in $\hat{H}^{\mathrm{RHF-CI}}$. The CI Hamiltonian in Eq.~(\ref{e}) is then used to construct the effective CIPT Hamiltonian for the low-energy states. In this procedure, the CI Hamiltonian is supplemented by a second-order correction obtained by summing over the high-energy states; see Ref.~\cite{CIPT} for details. Solving the resulting CIPT equations yields the seven-electron wave functions and the corresponding energies.
}

For each level, we calculate the Land\'{e} $g$ factor as the expectation value of the operator describing the interaction of the electrons with an external magnetic field (M1). The calculated $g$ factors are compared with both the experimental values and the nonrelativistic expression
\begin{equation}\label{e:gf}
g_{nr} = 1 + \frac{J(J+1)-L(L+1)+S(S+1)}{2J(J+1)}.
\end{equation}
Here $J$ is the total electronic angular momentum, while $L$ and $S$ are the total orbital and spin angular momenta, respectively.

In the relativistic calculations, $L$ and $S$ are not conserving quantum numbers and therefore are not uniquely defined. To compare the calculated $g$ factors with Eq.~(\ref{e:gf}), we assign values of $L$ and $S$ that provide the best agreement between the calculated and nonrelativistic $g$ factors. Knowledge of the $g$ factors is useful for identifying atomic states. Comparison with Eq.~(\ref{e:gf}) also facilitates the assignment of spectroscopic labels and the grouping of states into fine-structure multiplets.

\subsection{Calculation of hyperfine structure}

To calculate the hyperfine structure, we use the time-dependent Hartree-Fock (TDHF) method (see, e.g. ~\cite{DzuFlaSilSus87}), which is equivalent to the well-known random-phase approximation (RPA).
The RPA equations have a form:	
\begin{equation}\label{e:RPA}
	\left(\hat H^{\rm RHF}_c-\epsilon_c\right)\delta\psi_c=-\left(\hat f+\delta V^{f}_{\rm core}\right)\psi_c
\end{equation}
where $\hat f$ is an operator of  an external field (nuclear magnetic dipole or electric quadrupole fields).  
The index $c$ in (\ref{e:RPA}) labels core states, $\psi_c$ is a single-electron wave function of the state $c$ in the core, $\delta\psi_c$ is the correction to this wave function caused by an external field, and $\delta V^{f}_{\rm core}$ is the correction to the self-consistent RHF potential caused by the response of all core electrons.
Equation (\ref{e:RPA}) is solved self-consistently for all states in the core. As a result, an effective operator of the interaction of valence electrons with an external field is constructed as $\hat f + \delta V^{f}_{\rm core}$. The energy shift of a many-electron state $a$ is given by
\begin{equation} \label{e:de}
\delta \epsilon_a = \langle a | \sum_{i=1}^{N_v} \left(\hat f+\delta V^f_{\rm core} \right)_i | a\rangle.
\end{equation}

When the wave function for the valence electrons comes as a solution of the CI equation ($\Psi_a = \sum_i x_i\Phi_i$), Eq.~(\ref{e:de}) is reduced to
\begin{equation}\label{e:mex}
\delta \epsilon_a = \sum_{ij} x_i x_j \langle \Phi_i|\hat H^{\rm HFS}|\Phi_j \rangle,
\end{equation}
where $\hat H^{\rm HFS} =  \sum_{i=1}^{N_v} (\hat f+\delta V^f_{\rm core})_i$,
$\Phi_i$ are Slater determinants and $x_i$ are expansion coefficients.

Energy shift (\ref{e:de}) is used to calculate HFS constants $A$ and $B$ using textbook formulas
\begin{equation}
A_a = \frac{g_I \delta \epsilon_a^{(A)}}{\sqrt{J_a(J_a+1)(2J_a+1)}},
\label{e:Ahfs}
\end{equation}
and
\begin{equation}
B_a = -2Q \delta \epsilon_a^{(B)}\sqrt{\frac{J_a(2J_a-1)}{(2J_a+3)(2J_a+1)(J_a+1)}}. 
\label{e:Bhfs}
\end{equation}
Here $\delta \epsilon_a^{(A)}$ is the energy shift (\ref{e:de})  caused by the interaction of atomic electrons with the nuclear magnetic moment $\mu$, $g_I=\mu/I$, $I$ is nuclear spin; $\delta \epsilon_a^{(B)}$ is the energy shift (\ref{e:de}) caused by the interaction of atomic electrons with the nuclear electric quadrupole moment $Q$ ($Q$ in (\ref{e:Bhfs}) is measured in barns). 

\subsection{Breit and QED corrections}

\label{s:QED}

It is well known that the inclusion of Breit and QED corrections is essential for obtaining accurate and reliable results in calculations of heavy atoms (see, e.g.~\cite{Indelicato}). 
Their role is even more important for the hyperfine structure, since the dominant contribution originates from short distances, where relativistic effects are enhanced.

The proper inclusion of Breit and QED corrections in many-electron atoms is a challenging task because of many-body effects such as relaxation~\cite{relaxation} and configuration interaction (CI). The latter can be strongly enhanced when states belonging to different configurations are separated by small energy intervals. If these states have significantly different hyperfine-structure matrix elements, the corresponding physical states become highly sensitive to configuration mixing. Breit and QED corrections modify the energy separations, thereby affecting the mixing and, consequently, the hyperfine structure.

The inclusion of Breit interaction is relatively straightforward.
The two-particle inter-electron interaction operator contains Coulomb and Breit terms
\begin{eqnarray}
\hat h_2= \frac{1}{r_{ij}} - \frac{\bm{\alpha}_i\cdot\bm{\alpha}_j+ 
(\bm{\alpha}_i\cdot \mathbf{n}_i)(\bm{\alpha}_j\cdot\mathbf{n}_j)}{2r_{ij}},
\label{e:h2}
\end{eqnarray}
where $\bm{\alpha}_j$ is the Dirac matrix. Therefore, the Breit interaction can be treated on an equal footing with the Coulomb interaction by constructing the Breit potential and using the sum of the Coulomb and Breit potentials at all stages of the calculations, including the initial relativistic Hartree--Fock (RHF) calculations for the core, the generation of the basis set, and the CI calculations.

The inclusion of QED corrections is more complicated. There are two distinct contributions to the QED correction to the hyperfine structure. One is the correction to the hyperfine interaction operator, or equivalently to the single-electron matrix elements, analogous to the situation in hydrogen-like ions. The other is a many-body effect that influences the CI calculations in the same way as described above for the Breit interaction.

To account for the latter indirect effect, we use the radiative potential, which includes both vacuum-polarization and self-energy contributions~\cite{radpot}. This potential is added to the RHF potential at all stages of the CI  calculations, in the same way as the Breit interaction. By shifting the energies of the atomic states, the QED potential affects the configuration mixing coefficients and, consequently, the hyperfine structure.

However, the radiative potential does not provide sufficiently accurate corrections to the matrix elements of singular operators, such as the weak interaction and hyperfine interaction ~\cite{radpot,PhysRevLett.89.283002,Kuchiev_2003}. Therefore, the QED correction to the hyperfine interaction operator itself must be treated separately.
To do so, we use the results of the sophisticated calculations of the QED corrections to the HFS of the single-electron ions performed in 
Refs.~\cite{Blundell1997,Cheng2006}.
For the $s_{1/2}$ states we use the result of Ref.~\cite{Blundell1997} for the QED corrections to the $s_{1/2}$ states of the hydrogen-like ions.
The results are presented as ratios of the QED corrections to the single-particle HFS matrix elements for wide range of nuclear charge $Z$.  
Its dependence on $Z$ (see last column of Table~III in \cite{Blundell1997}) can be approximated by
\begin{equation}\label{e:qs}
Q_s(Z) = \frac{\alpha}{\pi}(-2.13 \times 10^{-4} Z^2 - 4.89 \times 10^{-2} Z + 0.318).
\end{equation}
Then, the inclusion of the QED corrections for any $Z$ can be done via correcting HFS matrix elements
\begin{equation}\label{e:Q}
\langle ns_{1/2}|\hat h_{\rm HFS} | ns_{1/2}\rangle \rightarrow (1+Q_s(Z))\langle ns_{1/2}|\hat h_{\rm HFS} | ns_{1/2}\rangle.
\end{equation}
 {Note that electron wave functions in this matrix element  do not include radiative correction produced by  $V_{\rm QED}$.} For $p_{1/2}$ states we use the results of Ref.~\cite{Cheng2006} (see last line of Table~II).
Again, the dependence on $Z$ is approximated by 
\begin{equation}\label{e:qp}
Q_p(Z) = \frac{\alpha}{\pi}(1.60 \times 10^{-5} Z^2 - 7.26 \times 10^{-3} Z + 0.206),
\end{equation}
and expressions for the matrix elements are corrected by expression similar to (\ref{e:Q}).

We neglect the QED corrections for the states other than $s_{1/2}$ or $p_{1/2}$, i.e. corresponding $Q$-factors are assumed to be equal to zero.
The values of $Q_s$ and $Q_p$ for atoms of current interest are presented in Table~\ref{t:Q}.

\begin{table}[ht]
\caption{The ratios (\ref{e:qs},\ref{e:qp}) for atoms of current interest. Numbers in square brackets stand for powers of ten} 
\label{t:Q}
\begin{ruledtabular}
\begin{tabular}{lr ccc}
\multicolumn{1}{c}{Atom}&
\multicolumn{1}{c}{$Z$}&
\multicolumn{1}{c}{$Q_s$}&
\multicolumn{1}{c}{$Q_p$}&
\multicolumn{1}{c}{$Q_p/Q_s$}\\
\hline
Xe  &  54 & -6.84[-3] & -3.30[-4] &  4.82[-2] \\
Rn  &  86 & -1.27[-2] & -7.03[-4] &  5.54[-2] \\
Og  & 118 & -1.95[-2] & -1.00[-3] &  5.12[-2] \\
\end{tabular}
\end{ruledtabular}
\end{table}

Note that all values of $Q_s$ and $Q_p$ are negative and maximum correction is -2\%.
{ In many-electron atoms the matrix elements of the HFS interaction for state $a$ are presented by Eq.~(\ref{e:mex}) in which the single-determinant
matrix element is a double sum over valence electrons
\begin{equation}\label{e:hfs}
\langle \Phi_i|\hat H^{\rm HFS}|\Phi_j \rangle = \sum_{i',j'}\langle i' | \hat h_{\rm HFS} | j'\rangle.
\end{equation}
}
$ \hat H_{\rm HFS} =  \sum_{i'} \hat h^{(i')}_{\rm HFS}$, summation goes over valence electrons. QED corrections are included by rescaling single-electron HFS matrix elements in the right-hand-side of (\ref{e:hfs}) using rescaling factors $Q$ from Table~\ref{t:Q} and Eq.~(\ref{e:Q}).
Note that the rescaling factors do not depend on the principal quantum number $n$, therefore they are applicable for both diagonal and off-diagonal
matrix elements.

\subsection{Enhancement of the Breit and QED effects in Og~II due to strong configuration mixing }

\label{s:mixing}

{To illustrate the effect of configuration mixing on the Breit and QED corrections to the hyperfine structure, we consider the  problem for  the lowest odd-parity states of Og II with total angular momentum $J=1/2$.} There is strong configuration mixing between the $7s^27p^48p$ and $7s^27p^5$ configurations:
\begin{equation}\label{e:x1}
X^{(1)}_{1/2}=7s^27p_{1/2}^27p_{3/2}^28p_{1/2}\,,
\end{equation}
\begin{equation}\label{e:x2}
X^{(2)}_{1/2}=7s^27p_{1/2}^27p_{3/2}^28p_{3/2}\,,
\end{equation}
and
\begin{equation}\label{e:y}
Y_{1/2}=7s^27p_{3/2}^47p_{1/2}.
\end{equation}
Note that the closed subshells $7s^2$, $7p_{1/2}^2$, and $7p_{3/2}^4$ do not contribute to the hyperfine structure. Recall also that electron wave functions with $j>1/2$ are suppressed in the vicinity of the nucleus and therefore make only small contributions to the hyperfine interaction. Consequently, the hyperfine constant $A_{\rm HFS}(X_{1/2})$ is determined mainly by the $8p_{1/2}$ component of the valence electron, whereas $A_{\rm HFS}(Y_{1/2})$ is dominated by the contribution of the $7p_{1/2}$ electron. {Numerical calculations gave single-particle HFS constants   $A_{\rm HFS}(7p_{1/2})$=187 GHz,   $A_{\rm HFS}(8p_{1/2})$=13.8  GHz,  and  $A_{\rm HFS}(8p_{3/2})$=0.30 GHz (assuming nuclear magnetic moment to nuclear spin ratio  $\mu/I$= 1 $\mu_0$,  where $\mu_0$ is nuclear magneton).  }
The large difference between $A_{\rm HFS}(X_{1/2})$ and $A_{\rm HFS}(Y_{1/2})$, together with the small energy separation between the interacting states, makes the hyperfine constants of the mixed states highly sensitive even to relatively small effects, such as the Breit and QED corrections.


\begin{table*}[ht]
\caption{The effect of Breit and QED corrections on the energies and magnetic
dipole HFS constants $A$ in case of strong configuration mixing. First three
odd states of Og~II with $J=1/2$ are considered. There are two strongly mixed
configurations: $X=7s^27p^48p$ and $Y=7s^27p^5$. Wave functions are written as
$a_1X^{(1)}_{1/2}+a_2X^{(2)}_{1/2}+bY_{1/2}$, where $X^{(1)}_{1/2}$, $X^{(2)}_{1/2}$ and $Y_{1/2}$ are given by (\ref{e:x1},\ref{e:x2},\ref{e:y}).
$\Delta$ is the difference of the current results
with the initial results with no Breit or QED included. {HFS constants are calculated assuming $\mu/I=1\,\mu_0$\,.}}
\label{t:Breit+QED}
\begin{ruledtabular}
\begin{tabular}{rrrrrrrr}

\multicolumn{1}{c}{$a_1^2$}&
\multicolumn{1}{c}{$a_2^2$}&
\multicolumn{1}{c}{$b^2$}&
\multicolumn{1}{c}{Other}&
\multicolumn{1}{c}{Energy}&
\multicolumn{1}{c}{$\Delta$}&
\multicolumn{1}{c}{$A$}&
\multicolumn{1}{c}{$\Delta$}\\

&&&&\multicolumn{1}{c}{cm$^{-1}$}&
\multicolumn{1}{c}{cm$^{-1}$}&
\multicolumn{1}{c}{GHz}&
\multicolumn{1}{c}{GHz}\\
\hline
\multicolumn{8}{c}{No Breit and no QED} \\
0.342 & 0.655 & 0        & 0.003 & 79625 & &   8.21 & \\
0.530 & 0.263 & 0.204 & 0.003 & 81843 & &  41.16 & \\                 
0.136 & 0.082 & 0.776 & 0.006 & 85116 & & 136.05 & \\
\multicolumn{8}{c}{With Breit but no QED} \\
0.356 & 0.641 & 0         & 0.003 & 79673 &   48 &   7.81 &  -0.40 \\
0.441 & 0.235 & 0.319 & 0.005 & 81637 & -212 &  61.94 &  20.78 \\     
0.209 & 0.125  & 0.661 & 0.005 & 84450 & -666 & 115.00 & -21.05 \\

\multicolumn{8}{c}{No Breit but with QED} \\
0.343 & 0.654 & 0        & 0.003 & 79643 &   18 &   8.08 &  -0.13 \\
0.525 & 0.263 & 0.212 & 0       & 81795 & -48 &  42.02&  0.86 \\   
0.140 & 0.083 & 0.777 & 0        & 85024 & -92 & 135.45& -0.60 \\

\multicolumn{8}{c}{With Breit and QED} \\
0.358 & 0.639 & 0        & 0.003       & 79670 &   45 &   7.84 &  -0.37 \\
0.432 & 0.231 & 0.332 & 0.005 & 81610 & -233 &  63.90 &  22.74 \\   
0.217 & 0.130 & 0.648 & 0.005 & 84395 & -721 & 112.17 & -23.88 \\

\end{tabular}
\end{ruledtabular}
\end{table*}

Table~\ref{t:Breit+QED} presents the calculated energies and magnetic-dipole hyperfine constants for the three lowest odd-parity states of Og~II with $J=1/2$. All three states can be approximately represented as mixtures of the three dominating configurations
\[
a_1 X_{1/2}^{(1)} +a_2  X_{1/2}^{(2)}+bY_{1/2}\,.
\]
{One can see that the lowest state is almost pure linear combination of $X_{1/2}$ states, 
\[
X_{1/2}=a_1 X_{1/2}^{(1)} +a_2  X_{1/2}^{(2)}\,,
\]
whereas the other two states are strong mixtures of $ X_{1/2}^{(1)}$, $ X_{1/2}^{(2)}$ and $Y_{1/2}$ states. }




When the Breit interaction is included (see Table~\ref{t:Breit+QED}), the energy separation between the second and third states decreases by only 14\%, whereas the hyperfine constant of the second state increases by 51\% and that of the third state decreases by 15\%.  Domination of the configuration mixing effect  follows from the fact that the sum of the corrections $\Delta$ to HFS for the second and third states is close to zero. A similar effect is observed when the QED corrections are included, although the changes are smaller: -12\% and +4\% for the hyperfine constants of the second and third states, respectively.

The last rows of Table~\ref{t:Breit+QED} present the final results obtained with the Breit and QED corrections included simultaneously. Note that the resulting correction differs significantly from the sum of the Breit and QED corrections calculated separately. This indicates that in the case of close levels  these corrections are sufficiently large to influence each other and therefore should be treated simultaneously.

\subsection{Calculations for Xe~II}

\begin{table*}[ht]
  \caption{Energy levels ($E$), $g$-factors and HFS constants ($A$ and $B$)
    of some excited states of Xe~II belonging to the $5s^25p^45d$ and $5s^25p^46p$ configurations.
    Comparison of theory with experiment. Magnetic dipole HFS constants $A$ are presented
    for $^{129}$Xe ($\mu = -0.7768\mu_0$, $I$=1/2~\cite{NIST_ASD}), electric quadrupole HFS
    constants $B$ are presented for $^{131}$Xe ($Q=-0.114b$~\cite{Stone2005}).}
\label{t:Xe}
\begin{ruledtabular}
\begin{tabular}{cc rc rc cc rr}
&&\multicolumn{2}{c}{Experiment~\cite{NIST_ASD}}&
\multicolumn{2}{c}{Calculation}&
\multicolumn{2}{c}{Experiment~\cite{Brostrom1996,Pawelec2011}}&
\multicolumn{2}{c}{Calculation}\\
\multicolumn{1}{c}{Conf.}&
\multicolumn{1}{c}{$J$}&
\multicolumn{1}{c}{$E$}&
\multicolumn{1}{c}{$g$}&
\multicolumn{1}{c}{$E$}&
\multicolumn{1}{c}{$g$}&
\multicolumn{1}{c}{$A$}&
\multicolumn{1}{c}{$B$}&
\multicolumn{1}{c}{$A$}&
\multicolumn{1}{c}{$B$}\\
&&\multicolumn{1}{c}{[cm$^{-1}$]}&&
\multicolumn{1}{c}{{[cm$^{-1}]$}}&&
\multicolumn{1}{c}{[MHz]}&
\multicolumn{1}{c}{[MHz]}&
\multicolumn{1}{c}{[MHz]}&
\multicolumn{1}{c}{[MHz]}\\

\hline

 $5d$ & 7/2 &  95438 & 1.42 &  96355 & 1.39 &  -502(4)  &  70(13)  &  -571 &   72 \\ 
 $5d$ & 7/2 & 101535 & 1.11 & 102280 & 1.14 &  -955(5)  & -150(50) & -1070 & -141 \\ 
 $6p$ & 5/2 & 111959 & 1.47 & 112781 & 1.48 & -1633(11) & -129(16) & -1648 & -130 \\ 
 $6p$ & 5/2 & 113512 & 1.28 & 114431 & 1.25 & -1387(9)  & -117(10) & -1404 & -115 \\ 
 $6p$ & 7/2 & 113705 & 1.40 & 114471 & 1.40 & -1164(5)  & -230(11) & -1215 & -257 \\ 
\end{tabular}
\end{ruledtabular}
\end{table*}

The Xe atom and its ions are well-studied systems. Their energy levels and $g$ factors are tabulated in the NIST database~\cite{NIST_ASD}, and the hyperfine structure of many excited states has been measured. This provides an opportunity to test our computational method and assess the accuracy that can be achieved.

Table~\ref{t:Xe} compares the calculated and experimental values of the energy levels, $g$ factors, and hyperfine constants for those states of Xe~II for which experimental hyperfine-structure data are available. We note that other theoretical calculations of the hyperfine structure of Xe~II have also been reported (see, e.g., Ref.~\cite{Paduch2000}). However, we do not discuss them here, since the primary purpose of this section is to demonstrate that the present approach accurately reproduces the available experimental data for Xe~II and therefore can be applied with confidence to Rn~II and Og~II.

The agreement between theory and experiment is generally good. It is at the level of about 1\% for the energies and approximately 10\% for the hyperfine constants. For the largest hyperfine constants, the agreement is even better, reaching about 1\% for some states. This is because small hyperfine constants are often the result of cancellations between different contributions, and such cancellations typically reduce the accuracy of the calculations. Therefore, for the purpose of extracting nuclear moments from hyperfine-structure measurements, one should focus on states with the largest hyperfine constants.

Based on the agreement obtained for Xe~II, it is reasonable to expect a similar level of accuracy for Rn~II and Og~II, namely, at least few percent for the energies and the largest hyperfine constants.
The accuracy may be lower for states affected by strong configuration mixing. The most prominent example, involving the mixing of the second and third odd-parity states with $J=1/2$, was discussed in Section~\ref{s:mixing}.

\section{Results of calculations for Rn~II and Og~II ions}.

\begin{table}[ht]
  \caption{Calculated energy levels ($E$), $g$-factors and HFS constants ($A$ and $B$) for ground and
    excited states of Rn~II. {HFS constants are calculated assuming $\mu/I=1\,\mu_0$ and $Q=1$ b.
    $N$ denotes the state number.}}
    \label{t:Rn}
\begin{ruledtabular}
\begin{tabular}{rlc rc rr}

\multicolumn{1}{c}{$N$}&
\multicolumn{1}{c}{Conf.}&
\multicolumn{1}{c}{State}&
\multicolumn{1}{c}{$E$}&
\multicolumn{1}{c}{$g$}&
\multicolumn{1}{c}{$A$}&
\multicolumn{1}{c}{$B$}\\
&&&\multicolumn{1}{c}{{[cm$^{-1}]$}}&&
\multicolumn{1}{c}{[MHz]}&
\multicolumn{1}{c}{[MHz]}\\
\hline
  1 & $6s^{2}6p^{5}$          & $^2$P$^o_{ 3/2}$ &        0 &  1.3333 &   1354 &  -3648 \\
  2 & $6s^{2}6p^{5}$          & $^2$P$^o_{ 1/2}$ &    29925\footnotemark[1] &  0.6667 &  24412 &      0 \\
  3 & $6s^{2}6p^{4}7s$        & $^4$D$_{ 5/2}$ &    75384 &  1.5112 &   3991 &   2093 \\ 
  4 & $6s^{2}6p^{4}7s$        & $^4$D$_{ 3/2}$ &    77450 &  1.2631 &   -917 &    636 \\  
  5 & $6s^{2}6p^{4}7s$        & $^2$S$_{ 1/2}$ &    85390 &  2.1803 &  10296 &      0 \\  
  6 & $6s^{2}6p^{4}6d$        & $^4$D$_{ 3/2}$ &    85841 &  1.1865 &   1202 &   -894 \\ 
  7 & $6s^{2}6p^{4}6d$        & $^6$F$_{ 5/2}$ &    85865 &  1.2985 &    797 &    -63 \\
  8 & $6s^{2}6p^{4}6d$        & $^4$D$_{ 7/2}$ &    86230 &  1.3326 &    475 &    340 \\
  9 & $6s^{2}6p^{4}6d$        & $^2$P$_{ 1/2}$ &    86798 &  0.7701 &   1848 &      0 \\  
 10 & $6s^{2}6p^{4}6d$        & $^2$F$_{ 7/2}$ &    89383 &  1.0955 &    859 &    744 \\
 11 & $6s^{2}6p^{4}6d$        & $^2$S$_{ 1/2}$ &    92325 &  1.9878 &   4035 &      0 \\ 
 12 & $6s^{2}6p^{4}6d$        & $^4$D$_{ 3/2}$ &    93880 &  1.2463 &    220 &   -380 \\ 
 13 & $6s^{2}6p^{4}7p$        & $^2$P$^o_{ 3/2}$ &    94997 &  1.5284 &   1226 &   1228 \\
 14 & $6s^{2}6p^{4}7p$        & $^8$G$^o_{ 5/2}$ &    95309 &  1.2743 &   1670 &   1861 \\
 15 & $6s^{2}6p^{4}6d$        & $^2$D$_{ 5/2}$ &    95773 &  1.1475 &    565 &      359 \\
 16 & $6s^{2}6p^{4}7p$        & $^4$D$^o_{ 5/2}$ &    99141 &  1.3467 &   1014 &    605 \\
 17 & $6s^{2}6p^{4}7p$        & $^2$S$^o_{ 1/2}$ &    99273 &  1.3597 &   2439 &      0 \\
 18 & $6s^{2}6p^{4}7p$        & $^8$G$^o_{ 7/2}$ &    99357 &  1.3623 &    939 &   2369 \\
 19 & $6s^{2}6p^{4}6d$        & $^2$D$_{ 3/2}$ &   100404 &  0.7878 &    401 &    600 \\   
 20 & $6s^{2}6p^{4}6d$        & $^2$D$_{ 5/2}$ &   101878 &  1.1898 &    -71 &   -512 \\
 21 & $6s^{2}6p^{4}7p$        & $^2$P$^o_{ 3/2}$ &   102198 &  1.3662 &   1221 &   -145 \\
 22 & $6s^{2}6p^{4}7s$        & $^2$P$_{ 1/2}$ &   107669 &  1.2868 & -10114 &      0 \\       
 23 & $6s^{2}6p^{4}7p$        & $^2$P$^o_{ 3/2}$ &   111411 &  1.3357 &    284 &    313 \\ 
 24 & $6s^{2}6p^{4}6d$        & $^4$F$_{ 7/2}$ &   117104 &  1.2811 &  -1973 &  -1723 \\
 25 & $6s^{2}6p^{4}6d$        & $^4$G$_{ 7/2}$ &   123187 &  0.9816 &   4514 &  -5519 \\
 26 & $6s^{2}6p^{4}7p$        & $^4$D$^o_{ 5/2}$ &   129424 &  1.3979 &  -1935 &  -1499 \\
 27 & $6s^{2}6p^{4}7p$        & $^4$F$^o_{ 5/2}$ &   131849 &  0.9994 &   6480 &  -6036 \\
\end{tabular}
\footnotetext[1]{NIST value is 30895.1 cm$^{-1}$~\cite{NIST_ASD}.} 
\end{ruledtabular}
\end{table}

\begin{table}[ht]
  \caption{Calculated energy levels ($E$), $g$-factors and HFS constants ($A$ and $B$) for ground and
    excited states of Og~II. {HFS constants are calculated assuming $\mu/I=1\,\mu_0$ and $Q=1$ b.}}
    \label{t:Og}
\begin{ruledtabular}
\begin{tabular}{rlc rc rr}

\multicolumn{1}{c}{$N$}&
\multicolumn{1}{c}{Conf.}&
\multicolumn{1}{c}{State}&
\multicolumn{1}{c}{$E$}&
\multicolumn{1}{c}{$g$}&
\multicolumn{1}{c}{$A$}&
\multicolumn{1}{c}{$B$}\\
&&&\multicolumn{1}{c}{{[cm$^{-1}]$}}&&
\multicolumn{1}{c}{[MHz]}&
\multicolumn{1}{c}{[MHz]}\\
\hline
  1 & $7s^{2}7p^{5}$          & $^2$P$^o_{ 3/2}$ &        0 &  1.3336 &  -2232 &  -6106 \\
  2 & $7s^{2}7p^{4}8s$        & $^4$D$_{ 5/2}$ &    47005 &  1.4817 &  11509 &   2779 \\  
  3 & $7s^{2}7p^{4}8s$        & $^4$D$_{ 3/2}$ &    50646 &  1.2199 & -13258 &   1106 \\ 
  4 & $7s^{2}7p^{4}8s$        & $^2$S$_{ 1/2}$ &    59086 &  2.0184 &  58551 &      0 \\  
  5 & $7s^{2}7p^{4}8p$        & $^2$P$^o_{ 3/2}$ &    68983 &  1.4865 &  -5972 &   1690 \\    
  6 & $7s^{2}7p^{4}7d$        & $^4$D$_{ 3/2}$ &    69128 &  1.1751 &  -1786 &  -1960 \\  
  7 & $7s^{2}7p^{4}8p$        & $^2$D$^o_{ 5/2}$ &    69151 &  1.2129 &   1214 &   2110 \\
  8 & $7s^{2}7p^{4}7d$        & $^8$G$_{ 5/2}$ &    69478 &  1.2532 &  -1586 &   -575 \\ 
  9 & $7s^{2}7p^{4}7d$        & $^2$P$_{ 1/2}$ &    69616 &  0.9880 &   -846 &      0 \\   
 10 & $7s^{2}7p^{4}7d$        & $^4$D$_{ 7/2}$ &    70030 &  1.2962 &  -1437 &    797 \\ 
 11 & $7s^{2}7p^{4}7d$       & $^8$H$_{ 7/2}$ &    71495 &  1.0845 &  -1199 &   3289 \\ 
 12 & $7s^{2}7p^{4}7d$        & $^2$S$_{ 1/2}$ &    74383 &  1.9048 &   1789 &      0 \\   
 13 & $7s^{2}7p^{4}7d$        & $^4$D$_{ 3/2}$ &    76256 &  1.2310 &   -687 &  -1015 \\ 
 14 & $7s^{2}7p^{4}7d$        & $^2$D$_{ 5/2}$ &    77784 &  1.1586 &   -792 &   1783 \\ 
 15 & $7s^{2}7p^{4}8p$        & $^2$P$^o_{ 1/2}$ &    78236 &  0.9122 &   7778 &      0 \\  
 16 & $7s^{2}7p^{4}8p$        & $^8$G$^o_{ 7/2}$ &    79986 &  1.3430 &  -1290 &   4071 \\
 17 & $7s^{2}7p^{4}8p$        & $^4$D$^o_{ 5/2}$ &    79991 &  1.3492 &  -1411 &   711 \\  
 18 & $7s^{2}7p^{4}8p$\footnotemark[1]        & $^2$P$^o_{ 1/2}$ &    81610 &  1.0031 &  63900 &      0 \\ %
 19 & $7s^{2}7p^{4}7d$        & $^2$D$_{ 3/2}$ &    82170 &  0.8403 &   -205 &   2081 \\  
 20 & $7s^{2}7p^{4}8p$        & $^2$P$^o_{ 3/2}$ &    82599 &  1.3471 &  -1897 &   -331 \\   
 21 & $7s^{2}7p^{4}7d$        & $^2$D$_{ 5/2}$ &    83878 &  1.1925 &   -283 &    810 \\  
 22 & $7s^{2}7p^{5}$\footnotemark[1]          & $^2$P$^o_{ 1/2}$ &    84395 &  0.7682 & 11217 &      0 \\ %
 23 & $7s^{2}7p^{4}8p$        & $^2$P$^o_{ 3/2}$ &    91232 &  1.3331 &     68 &    880 \\
 24 & $7s^{2}7p^{4}8d$        & $^2$S$_{ 1/2}$ &   107580 &  1.3709 &  -4037 &      0 \\    
 25 & $7s^{2}7p^{4}8d$        & $^2$F$_{ 7/2}$ &   107732 &  1.1867 &  -1490 &   2691 \\ 
 26 & $7s^{2}7p^{4}8d$        & $^4$F$_{ 7/2}$ &   108090 &  1.1927 &  -1163 &    759 \\  
\end{tabular}
\footnotetext[1]{States with strong configuration mixing.} 
\end{ruledtabular}
\end{table}

\begin{table}[ht]
  \caption{The experimental (for Xe~II, \cite{NIST_ASD}) and calculated $g$-factors for the $^4$D$_J$ fine structure multiplet of Xe~II, Rn~II and Og~II. 
  Comparison with non-relativistic values $g_{nr}$ (\ref{e:gf}).}
    \label{t:g}
\begin{ruledtabular}
\begin{tabular}{ccccc}

$J$ & $g_{nr}$ & Xe~II & Rn~II & Og~II \\
\hline
7/2 & 1.4286 & 1.42 & 1.3326 & 1.2962 \\ 
5/2 & 1.3714 & 1.56 & 1.5112 & 1.4817 \\
3/2 & 1.2000 & 1.38 & 1.2631 & 1.2199 \\
1/2 & 0.0000 & 0.50 & 0.7707 & 0.9818 \\
\end{tabular}
\end{ruledtabular}
\end{table}

The calculated energy levels, $g$ factors, and hyperfine-structure constants for Rn~II and Og~II are presented in Tables~\ref{t:Rn} and \ref{t:Og}. We retain the conventional non-relativistic spectroscopic labels to facilitate comparison between the light and heavy ions. However, these labels should be regarded as approximate because relativistic effects and configuration mixing become increasingly important with increasing nuclear charge.

For example, the fine-structure splitting of the ground-state doublet $^2P^{\rm o}_{3/2,1/2}$ increases more rapidly than $Z^2$. In Xe~II and Rn~II, the upper $^2P^{\rm o}_{1/2}$ level remains well separated from the rest of the spectrum (see Ref.~\cite{NIST_ASD} for Xe~II and Table~\ref{t:Rn} for Rn~II). In contrast, in Og~II this level lies in a region of high level density and is strongly mixed with states of other configurations (see Sec.~\ref{s:mixing} for a detailed discussion).

Another interesting example is the $^4$D$_J$ fine-structure multiplet. It can be identified in all three ions despite the strong relativistic effects in Rn~II and Og~II (see Table~\ref{t:g}). Although the calculated $g$-factors deviate significantly from their non-relativistic values, they remain sufficiently close to allow an unambiguous identification of the members of the multiplet.

The magnetic dipole ($A$) and electric quadrupole ($B$) hyperfine-structure constants for Rn~II and Og~II are presented in Tables~\ref{t:Rn} and \ref{t:Og} with the nuclear factors assumed to be equal to one. Specifically, the magnetic dipole constants are given without the factor $g_I=\mu/I$, where $\mu$ is the nuclear magnetic dipole moment and $I$ is the nuclear spin, while the electric quadrupole constants are given without the factor $Q$, the nuclear electric quadrupole moment.

To obtain the hyperfine constants for a particular radon isotope, the tabulated values should be multiplied by the corresponding nuclear moments taken from Ref.~\cite{Stone2005}. In contrast, the nuclear magnetic and electric quadrupole moments of oganesson isotopes are presently unknown. Therefore, the calculated hyperfine constants can serve as electronic-structure factors for extracting these nuclear moments from future hyperfine-structure measurements.

In Rn~II, the hyperfine structure can, in principle, be measured using the optical M1 transition between the ground state and the first excited state of the ground-state fine-structure doublet. The hyperfine splitting of the upper state is relatively large, which is advantageous for accurate measurements and interpretation of the spectra. Another possibility is to use electric-dipole (E1) transitions between excited states of opposite parity.

In Og~II, only the second option is available because the ground state is separated from the rest of the spectrum by a very large energy gap, exceeding the optical range. 
Of particular interest are the E1 transitions from the $7s^{2}7p^{4}8s,{}^{2}S_{1/2}$ state at $E=59086~\mathrm{cm}^{-1}$ to the $7s^{2}7p^{4}8p,{}^{2}P^{\rm o}_{1/2}$ 
state at $E=81610~\mathrm{cm}^{-1}$ and the $7s^{2}7p^{5},{}^{2}P^{\rm o}_{1/2}$ state at $E=84395~\mathrm{cm}^{-1}$. 
The hyperfine structure is large for all three states, making these transitions particularly suitable for spectroscopic studies.

The large hyperfine constants can be understood from the electronic structure of the states. In relativistic notation, the lower state is predominantly the $7s^27p_{3/2}^48s$ configuration. Since the closed $7s^2$ and $7p_{3/2}^4$ subshells do not contribute to the hyperfine interaction, the hyperfine structure is dominated by the $8s_{1/2}$ electron. This contribution is large because $s_{1/2}$ (and, to a lesser extent, $p_{1/2}$) orbitals have a significant probability density inside the nucleus. Similarly, the upper state is predominantly the $7s^27p_{3/2}^47p_{1/2}$ configuration, and its hyperfine structure is dominated by the $7p_{1/2}$ electron.

This transition is particularly suitable for determining the magnetic dipole hyperfine constant ($A$). On the other hand, it cannot be used for measuring the electric quadrupole constant ($B$), because both states have $J=1/2$, for which the electric quadrupole hyperfine interaction vanishes.

\section{Conclusion}

We have performed calculations of the energy levels, Land\'{e} $g$ factors, and hyperfine-structure constants for the Xe~II, Rn~II, and Og~II ions using the configuration interaction with perturbation theory (CIPT) method. The calculated energies, $g$ factors, and hyperfine constants for Xe~II are in good agreement with the available experimental data, providing confidence in the reliability of the approach when applied to the heavier ions Rn~II and Og~II.

Special attention has been paid to the treatment of Breit and quantum electrodynamic (QED) corrections. We have demonstrated that, in the presence of strong configuration mixing, relatively small Breit and QED corrections to the energies can lead to much larger changes in the hyperfine-structure constants. This enhancement is particularly pronounced when the mixing occurs between states with very different hyperfine structures and small energy separations. Under such conditions, the Breit and QED corrections become strongly nonadditive and must be included simultaneously.

The calculated spectra reveal a significant increase in relativistic effects along the sequence Xe~II--Rn~II--Og~II. Nevertheless, the conventional spectroscopic classification remains useful for identifying many low-lying states and fine-structure multiplets. For Og~II, strong configuration mixing substantially modifies the structure of some low-lying levels and plays a crucial role in determining their hyperfine properties.

The present calculations provide predictions for the energy levels and hyperfine structure of Rn~II and Og~II, for which little or no spectroscopic information is currently available. We identify several transitions that appear particularly promising for future measurements. The calculated hyperfine constants may be combined with experimental data to extract nuclear magnetic dipole and electric quadrupole moments. In particular, the results for Og~II may prove useful for determining nuclear moments of oganesson isotopes from future  measurements, for which no such information is presently available.

Overall, the present work provides a comprehensive description of the low-lying spectra and hyperfine structure of Rn~II and Og~II and highlights the importance of relativistic, correlation, Breit, and QED effects in heavy open-shell ions.

\section{Acknowledgements}
This work was supported by the Australian Research Council.

%

\end{document}